\documentclass[conference]{IEEEtran}
\ifCLASSINFOpdf
\else
\fi

\usepackage{amssymb,amsfonts,amsmath,amsthm,amscd,dsfont,mathrsfs}
\usepackage{graphicx,float,psfrag,pstcol,epsfig,color,subfigure}
\usepackage{psfrag}
\usepackage{pstcol}
\usepackage{import}

\newcommand{\cX}{{\cal X}}
\newcommand{\cY}{{\cal Y}}
\newcommand{\cU}{{\cal U}}

\theoremstyle{definition}
\theoremstyle{plain}
\newtheorem{thm}{Theorem}%[section]
\theoremstyle{plain}
\theoremstyle{plain}
\newtheorem{lemma}{Lemma}%[section]
\theoremstyle{plain}
\theoremstyle{plain}
\theoremstyle{plain}
\theoremstyle{remark}
\begin{document}

\title{Linear Information Coupling Problems}

\author{
\IEEEauthorblockN{Shao-Lun Huang}
\IEEEauthorblockA{Department of EECS\\
Massachusetts Institute of Technology\\
Cambridge, MA 02139-4307\\
Email: shaolun@mit.edu}
\and
\IEEEauthorblockN{Lizhong Zheng}
\IEEEauthorblockA{Department of EECS\\
Massachusetts Institute of Technology\\
Cambridge, MA 02139-4307\\
Email: lizhong@mit.edu}
}

\maketitle

\begin{abstract}

Many network information theory problems face the similar difficulty of single letterization. We argue that this is due to the lack of a geometric structure on the space of probability distribution. In this paper, we develop such a structure by assuming that the distributions of interest are close to each other. Under this assumption, the K-L divergence is reduced to the squared Euclidean metric in an Euclidean space. Moreover, we construct the notion of coordinate and inner product, which will facilitate solving communication problems. We will also present the application of this approach to the point-to-point channel and the general broadcast channel, which demonstrates how our technique simplifies information theory problems.

\end{abstract}

\IEEEpeerreviewmaketitle

\section{Introduction}

Since Shannon introduced the notion of capacity sixty years ago, finding the capacity of channels and networks are core problems in information theory. The analyzation of capacity answers the problem that how many bits can be transmitted through a communication network, and also provides many insights in engineer problems \cite{TJ91}. However, for general problems, there is no systematic way to obtain optimal single-letter solutions. By cleverly picking auxiliary random variables, we sometimes can prove the constructed single-letter solutions are optimal for some problems. But, when this fails, we cannot tell whether it is because we have not tried hard enough or the problem itself does not have an optimal single-letter solution.

The difficulty of obtaining optimal single letter solutions comes from the fact that, most of the information theoretical quantities, such as entropy, mutual information, and error exponents, are all special cases of the Kullback-Leibler (K-L) divergence. The K-L divergence is a measure of distance between two probability distributions. However, in multi-terminal communication problems, there are multiple input and output distributions, and we usually need to deal with problems in high dimensional probability spaces. In these cases, describing problems only with the distance measure is cumbersome, and solving these information theory problems turns out to be extremely hard even with numerical aids. Therefore, we need more geometric structures to describe the problems, such as inner products. This is however difficult, as the K-L divergence between two distributions, $D(P\|Q)$, is not symmetric between $P$ and $Q$, and the K-L divergence is in general not a valid metric. Thus, the space of probability distributions is not a linear vector space but a manifold \cite{SH00}, when the K-L divergence behaves as the distance measure.

In this paper, we present an approach \cite{SL08} to simplify the problems with the assumption that the distributions of interest are close to each other. With this assumption, the manifold formed by distributions can be approximated by a tangent plane, which is an Euclidean space. Moreover, the K-L divergence will behave as the squared Euclidean metric between distributions in this Euclidean space. Therefore, we obtain the notion of coordinate, inner product, and orthogonality in a linear metric space, to describe information theory problems. Moreover, we will demonstrate in the rest sections that, the linear structure constructed from out local approximation will transfer information theory problems to linear algebra problems, which can be solved easily. In particular, we show that a systematic approach can be used to solve the single-letterization problems. We apply this to the general broadcast channel, and obtain new insights of the optimality of the existing solutions \cite{T75}-\cite{K79}

The rest of this paper is organized as follows. We introduce the notion of local approximation in section \ref{sec:local}, and show that the K-L divergence can be approximated as the squared Euclidean metric. In section \ref{sec:p2p} and \ref{sec:bc}, we present the application of our local approximation to the point-to-point channel and the general broadcast channel, respectively. We will illustrate how the information theory problems become simple linear algebra problems when applying our technique.

%\hfill mds
 
%\hfill January 11, 2007

\section{The Local Approximation} \label{sec:local}

The key step of our approach is to use a local approximation of the Kullback-Leibler (K-L) divergence. Let $P$ and $Q$ be two distributions over the same alphabet $\cX$. We assume that $Q(x) = P(x) + \epsilon J(x)$, for some small value $\epsilon$, then the K-L divergence can be written, with second order Taylor expansion, as
\begin{align}
\notag
D(P \| Q) 
&= -\sum_{x} P(x) \log \frac{Q(x)}{P(x)} \\ \notag
&= -\sum_{x} P(x) \log \left( 1 + \epsilon \cdot \frac{J(x)}{P(x)} \right) \\ \notag
&= \frac{1}{2} \epsilon^2 \cdot \sum_{x} \frac{1}{P(x)} J^2(x) + o(\epsilon^2).
\end{align}
We denote $\sum_{x} J^2(x)/P(x)$ as $\| J \|^2_P$, which is the weighted norm square of the perturbation vector $J$. It is easy to verify here that replacing the weight in this norm by $Q$ only results in a $o(\epsilon^2)$ difference. That is, up to the first order approximation, the weights in the norm simply indicate the neighborhood of distributions where the divergence is computed. As a consequence, $D(P \| Q)$ and $D(Q \| P)$ are considered as equal up to the first order approximation.

For convenience, we define the weighted perturbation vector as
\begin{equation}\notag
L(x) \triangleq \frac{1}{\sqrt{P(x)}} J(x), \ \ \forall x,
\end{equation}
or in vector form $L \triangleq \left[ \sqrt{P}^{-1} \right] J$, where $\left[ \sqrt{P}^{-1} \right]$ represents the diagonal matrix with entries $\left\{ \sqrt{P(x)}^{-1}, \ x \in \cX \right\}$.	This	allows	us	to	write $\| J \|^2_P = \| L \|^2$, where the last norm is simply the Euclidean norm.

With this definition of the norm on the perturbations of distributions, we can generalize to define the corresponding notion of inner products. Let $Q_i (x) = P (x) + \epsilon \cdot J_i (x)$, $\forall x, i = 1, 2$, we can define
\begin{equation}\notag
\langle J_1 , J_2 \rangle_P \triangleq \sum_{x} \frac{1}{P(x)} J_1 (x) J_2 (x) = \langle L_1 , L_2 \rangle,
\end{equation}
where $L_i = \left[ \sqrt{P}^{-1} \right] J_i$, for $i = 1, 2$. From this, notions of orthogonal perturbations and projections can be similarly defined. The point here is that we can view a neighborhood of distributions as a linear metric space, and define notions of orthonormal basis and coordinates on it.

\section{The Point to Point Channel} \label{sec:p2p}

We start by using this local geometric structure to study the point-to-point channels to demonstrate the new insights we can obtain from this approach, even on a well-understood problem. It is well-known that the capacity problem is
\begin{align} \label{eq:cap}
\max_{P_X} I(X;Y),
\end{align}
but this is in fact a single letter solution of the coding problem
\begin{align} \label{eq:cap2}
\max_{U \rightarrow X^n \rightarrow Y^n} \frac{1}{n} I(U;Y^n),
\end{align}
for some discrete random variable $U$, such that $U \rightarrow X^n \rightarrow Y^n$ forms a Markov chain.

Now, to apply the local approximations, instead of solving \eqref{eq:cap2}. we study a slightly different problem
\begin{equation}\label{eq:LICP_multi-letter}
\max_{U \rightarrow X^n \rightarrow Y^n : \frac{1}{n} I(U;X^n) \leq \frac{1}{2} \epsilon^2} \frac{1}{n} I(U;Y^n).
\end{equation}
We call the problem \eqref{eq:LICP_multi-letter} as the \emph{linear information coupling problem}. The only difference between \eqref{eq:cap2} and \eqref{eq:LICP_multi-letter} lies in the constraint $\frac{1}{n} I(U;X^n) \leq \frac{1}{2} \epsilon^2$ on \eqref{eq:LICP_multi-letter}. That is, instead of trying to find how many bits in total that we can send through the given channel, we ask the question of how efficiently we can send a thin layer of information through this channel. One advantage of \eqref{eq:LICP_multi-letter} is that it allows easy single letterization as we will demonstrate in the following. In fact, the step of single-letterization, namely, form \eqref{eq:cap2} to \eqref{eq:cap}, is the difficult step of most network problems. For these problems, the approach we used for the point-to-point problem can not be applied. What we will show in the rest of this section is that there is an alternative approach to do the well-known steps \cite{TJ91} to go from \eqref{eq:cap2} to \eqref{eq:cap}, and this new approach based on the geometric structures can be applied to more general problems. For simplicity, in this paper, we assume that the marginal distribution $P_{X^n}$ is given, and is an i.i.d. distribution over the $n$ letters\footnote{This assumption can be proved to be ``without loss of the optimality'' for some cases \cite{SL08}. In general, it requires a separate optimization, which is not the main issue addressed in this paper. To that end, we also assume that the given marginal $P_{X^n}$has strictly positive entries.}, so that we can focus on finding $U$ and the conditional distribution $P_{X^n|U}$ optimizing \eqref{eq:LICP_multi-letter}. 

First, we solve the single-letter version, namely $n=1$, of this problem. Observing that we can write the constraint as
\begin{equation}\notag
I(U;X) = \sum_{u} P_U (u) \cdot D(P_{X|U} (\cdot | u) \| P_X ) \leq \frac{1}{2} \epsilon^2.
\end{equation}
This implies that for each value of $u$, the conditional distribution $P_{X|U=u}$ is a local perturbation from $P_X$, that is, $P_{X|U=u} = P_X + \epsilon \cdot J_u$. %Again,we use the notation that $L = [\sqrt{P}^{-1}] J$.

Next, using the notation that $L_u = \left[ \sqrt{P}^{-1} \right] J_u$, for each value of $u$, we observe that
\begin{align}
\notag
P_{Y|U=u} 
&= W P_{X|U=u} \\ \notag
&= WP_X + \epsilon \cdot W J_u \\ \notag
&= P_Y + \epsilon \cdot W [\sqrt{P_X}] L_u,
\end{align}
where the channel applied to an input distribution is simply viewed as the channel matrix $W$, of dimension $| \cY | \times | \cX |$, multiplying the input distribution as a vector. At this point, we have reduced both the spaces of input and output distributions as linear spaces, and the channel acts as a linear transform between these two spaces. The information coupling problem can be rewritten as, ignoring the $o(\epsilon^2)$ terms:
\begin{align*}
\max. \ &\sum_{u} P_U(u) \cdot \| W J_u \|^2_{P_Y}, \\ 
\mbox{subject to:} \ &\sum_{u} P_U(u) \cdot \| J_u \|^2_{P_X} = 1,
\end{align*}
or equivalently in terms of Euclidean norms,
\begin{align*}
\max. \ &\sum_{u} P_U(u) \cdot \| \left[\sqrt{P_Y}^{-1} \right] W \left[\sqrt{P_X} \right] \cdot L_u \|^2, \\
\mbox{subject to:} \ &\sum_{u} P_U(u) \cdot \| L_u \|^2 = 1.
\end{align*}

\begin{figure}
\centering 
\subfigure[]{
\includegraphics{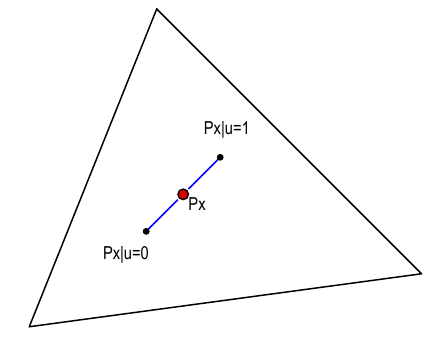} 
\label{fig:Perturb}
}
\subfigure[]{
\includegraphics{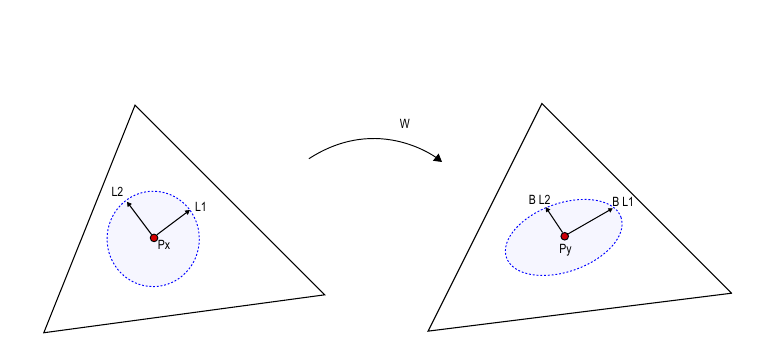}
\label{fig:Channel_Map}
}
\caption{(a) Choice of $P_U$ and $P_{X|U}$ to maintain the marginal $P_X$. (b) Divergence Transition Map as a linear map between two spaces, with right and left singular vectors as orthonormal bases.}
\end{figure}

This problem of linear algebra is simple. We need to find the joint distribution $U \rightarrow X \rightarrow Y$ by specifying the $P_U$ and the perturbations $J_u$ for each value of $u$, such that the marginal constraint on $P_X$ is met, and also these perturbations are the most visible at the $Y$ end, in the sense that multiplied by the channel matrix, $W J_u$'s have large norms. This can be readily solved by setting the weighted perturbation vectors $L_u$'s to be along the input (right) singular vectors of the matrix
$B \triangleq \left[\sqrt{P_Y}^{-1} \right] W \left[\sqrt{P_X} \right]$ with large singular values. Moreover, the choice of $P_U$ has no effect in the optimization, and might be taken as binary uniform for simplicity. This is illustrated in Figure \ref{fig:Perturb}.

We call the matrix $B$ the \emph{divergence transition matrix} (DTM). It maps divergence in the space of input distributions to that of the output distributions. The singular value decomposition (SVD) structure of this linear map has a critical role of our analysis. It can be shown that the largest singular value of $B$ is $1$, corresponding to an input singular vector $\left[ \sqrt{P_X}, x \in \cX \right]^T$, which is orthogonal to the simplex of probability distributions. This is not a valid choice for perturbation vectors. However, all vectors orthogonal to this vector, or equivalently, all linear combinations of other singular vectors are valid choices of the perturbation vectors $L_u$. Thus, the optimum of the above problem is achieved by setting $L_u$ to be along the singular vector with the second largest singular value.

This can be visualized as in Figure \ref{fig:Channel_Map}, the orthonormal bases for the input and output spaces, respectively, according to the right and left singular vectors of $B$. The key point here is that while $I(U;X)$ measures how many bits of information is modulated in $X$, depending on how they are modulated, in terms of which direction the corresponding perturbation vector is, these bits have different levels of visibility at the $Y$ end. This is a quantitative way to show why viewing a channel as a bit-pipe carrying uniform bits is a bad idea.

Moreover, recalling that the data processing inequality tells that, from the Markov chain $U \rightarrow X \rightarrow Y$, the mutual informations have the relation $I(U;X) \geq I(U;Y)$. Let us assume that the second largest singular value of $B$ is $\sigma \leq 1$, then the above derivations imply that $\sigma^2 \cdot I(U;X) \geq  I(U;Y)$. Thus, we actually come up with a stronger result than the data processing inequality, and the equality can be achieved by setting the perturbation vector to be along the right singular vector of $B$, with the second largest singular value.

The most important feature of the linear information coupling problem is that the single-letterization (\ref{eq:LICP_multi-letter}) is simple. To illustrate the idea, we consider a 2-letter version of the point-to-point channel:
\begin{equation}\label{eq:LICP_single-leterization}
\max_{U \rightarrow X^2 \rightarrow Y^2 : \frac{1}{2} I( U ; X^2 ) \leq  \frac{1}{2} \epsilon^2} \frac{1}{2} I( U ; Y^2 ).
\end{equation}
Let $P_X$, $P_Y$, $W$, and $B$ be the input and output distributions, channel matrix, and the DTM, respectively for the single letter version of the problem. Then, the 2-letter problem has $P^{(2)}_X = P_X \otimes P_X$, $P^{(2)}_Y = P_Y \otimes P_Y$, and $W^{(2)} = W \otimes W$, where $\otimes$ denotes the Kronecker product. As a result, the new DTM is $B^{(2)} = B \otimes B$. We have the following lemma on the singular values and vectors of $B^{(2)}$. %It is easy to verify that if $\mu_i, \mu_j$ are the singular values of $B$, corresponding to singular vectors $v_i, v_j$, then $\mu_i \mu_j$ is a singular value of $B^{(2)}$, with singular vector $v_i \otimes v_j$.

\begin{lemma}
Let $v_i$ and $v_j$ denote two singular vectors of $B$ with singular values $\mu_i$ and $\mu_j$. Then $v_i \otimes v_j$ is a singular vector of $B^{(2)}$ and its singular value is $\mu_i \mu_j$.
\end{lemma}

Now, recall that the largest singular value of $B$ is $\mu_0 = 1$, with the singular vector $v_0 = \left[ \sqrt{P_X}, x \in \cX \right]^T$, which corresponds to the direction orthogonal to the distribution simplex. This implies that the largest singular value of $B^{(2)}$ is also 1, again corresponds to the direction that is orthogonal to all valid choices of the perturbation vectors.

The second largest singular value of $B^{(2)}$ is a tie between $\mu_0 \mu_1$ and $\mu_1 \mu_0$, with singular vectors $v_0 \otimes v_1$ and $v_1 \otimes v_0$. The optimal solution of (\ref{eq:LICP_single-leterization}) is thus to set the perturbation vectors to be along these two vectors. This can be written as
\begin{align}
\notag
&P_{X^2|U=u} \\ \notag
=& P_X \otimes P_X + \left[\sqrt{P_X \otimes P_X} \right] \cdot \left( \epsilon v_0 \otimes v_1 + \epsilon' v_1 \otimes v_0 \right) \\ \notag
%=& P_X \otimes P_X + \epsilon \cdot \left( \left[ \sqrt{P_X} \right] v_0 \otimes \left[\sqrt{P_X}\right] v_1 \right) + \epsilon' \cdot \left( \left[\sqrt{P_X}\right] v_1 \otimes \left[\sqrt{P_X}\right] v_0 \right) \\ \notag
%&+ \epsilon \cdot \epsilon' \cdot \left( \left[\sqrt{P_X}\right] v_1 \otimes \left[\sqrt{P_X}\right] v_1 \right) + O(\epsilon^2) \\ \label{eq:aa}
=& \left( P_X + \epsilon' \left[\sqrt{P_X}\right] v_1 \right) \otimes \left( P_X + \epsilon \left[\sqrt{P_X}\right] v_1 \right) + O(\epsilon^2).
\end{align}
Here, we use the fact that $v_0 = \left[\sqrt{P_X}, x \in \cX \right]^T$. This means that the optimal conditional distribution $P_{X^2|U=u}$ for any $u$ has the product form, up to the first order approximation. With a simple time-sharing argument, it is easy to see that we can indeed set $\epsilon = \epsilon'$, that is, pick this conditional distribution to be i.i.d. over the two symbols, to achieve the optimum.

The simplicity of this proof of the optimality of the single letter solutions is astonishing. All we have used is the fact that the singular vector of $B^{(2)}$ corresponding to the second largest singular value has a special form. A distribution in the neighborhood of $P_X \otimes P_X$ is a product distribution if and only if it can be written as a perturbation from $P_X \otimes P_X$, along the subspace spanned by vectors $v_0 \otimes v_i$ and $v_j \otimes v_0$, in the form of $v_0 \otimes v + v' \otimes v_0$, for some $v$ and $v'$.
%One can visualize the space of $2$-letter joint distributions $P_{X^2}$ around the i.i.d. distribution $P_X \otimes P_X$ as perturbations from it, so there is a subspace corresponds to product distributions, with perturbation vectors of the form $v_0 \otimes v $ and $v \otimes v_0$ for some $v$. 
Thus, all we need to do is to find the eigen-structure of the $B$-matrix, and verify if the optimal solutions have this form. This procedure is used in more general problems.

One way to explain why the local approximation is useful is as follows. In general, tradeoff between multiple K-L divergence (mutual information) is a non-convex problem. Thus, finding global optimum for such problems is in general intrinsically intractable and extremely hard. In contrast, with our local approximation, the K-L divergence becomes a quadratic function. Now, the tradeoff between quadratic functions remains quadratic. Effectively, our approach focus on verifying the local optimality of the quadratic solutions, which is a natural thing to do, since the overall problem is not convex.

%This is not a coincidence. We will demonstrate in section \label{sec:bc} that in the general broadcast channel, the similar structure can be proved and used for single-letterization or finite-letterization. We would like to emphasize that the advantage of our approach is that it does not require any constructive proving technique. For any given problem, one can follow essentially the same procedure to find out the SVD structure of the corresponding DTM. The result either gives a proof of the optimality of the single letter solutions or disproves it without any ambiguity. This is clearly a very desirable feature in a field with a large number of open conjectures.

\section{The General Broadcast Channel} \label{sec:bc}

Let us now apply our local approximation approach to the general broadcast channel. A general broadcast channel with input $X \in \cX$, and outputs $Y_1 \in \cY_1$, $Y_2 \in \cY_2$, is specified by the memoryless channel matrices $W_1$ and $W_2$. These channel matrices specify the conditional distributions of the output signals at two users, $1$ and $2$ as $W_i ( y_i | x ) = P_{Y_i|X} ( y_i | x )$, for $i = 1,2$. Let $M_1$, $M_2$, and $M_0$ be the two private messages and the common message, with rate $R_1$, and $R_2$, and $R_0$, respectively. The multi-letter capacity region can be written as
\begin{equation} \notag
\left\{
\begin{array}{clr}
R_0 &\leq \frac{1}{n} \min \{ I(U ; Y^n_1) , I(U ; Y^n_2) \}, \\ 
R_1 &\leq \frac{1}{n} I(V_1 ; Y^n_1), \\
R_2 &\leq \frac{1}{n} I(V_2 ; Y^n_2),
\end{array} \right.
\end{equation}
for some mutually independent random variables $U $, $V_1 $, and $V_2$, such that $( U , V_1 , V_2 ) \rightarrow X^n \rightarrow (Y^n_1, Y^n_2)$ forms a Markov chains. 

The linear information coupling problems of the private messages, given that the common message is decoded, are essentially the same as the point-to-point channel case. Thus, we only need to focus on the linear information coupling problem of the common message:
%To apply our approach, we consider the following linear information coupling problems for the general broadcast channel:
\begin{align}
\label{eq:BC_single_public}
\max. & \ \frac{1}{n} \min \left\{ I(U ; Y^n_1) , I(U ; Y^n_2) \right\}, \\ \notag
\mbox{subject to:} & \ U \rightarrow X^n \rightarrow (Y^n_1, Y^n_2) : \frac{1}{n} I(U ; X^n) \leq \frac{1}{2} \epsilon^2
%\max_{U_1 \rightarrow X^n \rightarrow Y^n_1 : \frac{1}{n} I(U_1 ; X^n) \leq \frac{1}{2} \epsilon^2_1} &\frac{1}{n} I(U_1 ; Y^n_1), \\ \label{eq:BC_single_private2}
%\max_{U_2 \rightarrow X^n \rightarrow Y^n_2 : \frac{1}{n} I(U_2 ; X^n) \leq \frac{1}{2} \epsilon^2_2} &\frac{1}{n} I(U_2 ; Y^n_2).
\end{align}

%Like in the point-to-point case, we assume that the marginal distribution of $X^n$ is i.i.d. $P_X$, hence the marginal distributions on the two outputs are also i.i.d., $P_{Y_1}$, $P_{Y_2}$, respectively. We also assume that $P_{U_i}$ are binary equi-probably distributed, and write the conditional distributions as perturbations from the marginal $P_{X^n|U_i=u_i} = P_X^n + \epsilon_i \cdot J_{u_i} $, for $i = 0,1,2$.

%Then, the linear information coupling problems (\ref{eq:BC_single_private1}) and (\ref{eq:BC_single_private2}), for private messages $M_1$ and $M_2$, have exactly the same form as (\ref{eq:LICP_multi-letter}). Thus, by defining the single-letter DTM's $B_i \triangleq \left[\sqrt{P_{Y_i}}^{-1}\right] W_i \left[\sqrt{P_X}\right]$, for $i = 1,2$, we can solve (\ref{eq:BC_single_private1}) and (\ref{eq:BC_single_private2}) with the same routine as (\ref{eq:LICP_multi-letter}), and the single-letter solutions are optimal. 

%The optimization problem (\ref{eq:BC_single_public}), however, turns out to be very different from the other two. 
The core problem we want to address here is that whether or not the single-letter solutions are optimal for (\ref{eq:BC_single_public}). To do this, suppose that $P_{X^n|U=u} = P_X^{(n)} + \epsilon \cdot J_{u} $. Define the DTMs $B_i \triangleq \left[\sqrt{P_{Y_i}}^{-1}\right] W_i \left[\sqrt{P_X}\right]$, for $i = 1,2$, and the scaled perturbation $L_{u} = \left[\sqrt{P^{(n)}_{X}}^{-1}\right]J_{u}$, the problem then becomes
\begin{align}
\label{eq:BC_multi_public}
\max_{L_{u} : \| L_{u} \|^2 = 1}  \min \left\{ \| B_1^{(n)} L_{u}  \|^2 , \| B_2^{(n)} L_{u}  \|^2 \right\},
\end{align}
where $B_i^{(n)}$ is the $n^{th}$ Kronecker product of the single-letter DTM $B_i$, for $i = 1,2$.

Different from the point-to-point problem, we need to choose the perturbation vectors $L_{u}$ to have large images simultaneously through two different linear systems. In general, the tradeoff between two SVD structures can be rather messy problems. However, in this problem, for both $i = 1, 2$, $B_i^{(n)}$ have the special structure of being the Kronecker product of the single letter DTMÕs. Furthermore, both $B_1$ and $B_2$ have the largest singular value of $1$, corresponding to the same singular vector $\phi_0 = \left[\sqrt{P_X}, x \in \cX \right]^T$, although the rest of their SVD structures are not specified. The following theory characterizes the optimality of single-letter and finite-letter solutions for the general cases.
\begin{thm} \label{thm:BC}
Suppose that $B_i$ are DTM's for some DMC and input/output distributions, for $i = 1,2, \ldots , k$, then the linear information coupling problem 
\begin{align} \label{eq:k-BC}
\max_{L_{u} : \| L_{u} \|^2 = 1}  \min_{1 \leq i \leq k} \left\{ \| B_i^{(n)} L_{u}  \|^2 \right\},
\end{align}
has optimal single letter solutions for the case with $2$ receivers. In general, when there are $k > 2$ receivers, single letter solutions can not be optimal, when the cardinality $|\cU|$ is bounded by some function of $|\cX|$. However, there still exists $k$-letter solutions that are optimal.
\end{thm}
While we will not present the full proof of this result in this paper, it worth pointing out how conceptually straightforward it is. We can write the right singular vectors of the two DTMÕs, $B_1$ and $B_2$, as $\phi_0, \phi_1, \ldots , \phi_{n - 1}$ and $\varphi_0 , \varphi_1, \ldots ,\varphi_{n-1}$. The only structure we have is that $\phi_0 = \varphi_0 = \left[ \sqrt{P_X (x)}, x \in X \right]^T$, both correspond to the largest singular value of 1. For other vectors, the relation between the two bases can be written as a unitary matrix $\Psi$, with $\phi_i = \sum_j \Psi_{ij} \varphi_{j}$. Now, we can define an orthonormal basis for the space of multi-letter distributions on $X^n$. For example, with 2-letter distributions, we can use $\phi_i \otimes \phi_j,i,j \in \left\{0,1,...,n - 1\right\}$ and $(i,j) \neq (0,0)$. Note that any $L_u$ can be written as $L_u = \sum_{{i,j} \neq (0,0)} \alpha_{ij} \phi_i \otimes \phi_j$. If a perturbation vector $L_{u}$ has any non-zero component along $\phi_i \otimes \phi_j$ , with $i, j \neq 0$, we can always move this component to either $\phi_i \otimes \phi_0$ or $\phi_0 \otimes \phi_j$ to have, say, $\phi_i  \otimes \phi_0 = \sum_j \Psi_{ij} \varphi_j \otimes \varphi_0$. This results in larger norms of the output vectors through both channels. As a result, the optimizer of \eqref{eq:k-BC} can only have components on the vectors $\phi_0 \otimes \phi_j$ and $\phi_i \otimes \phi_0$. This means that the resulting conditional distribution must be product distributions, i.e. $P_{X^n|U=u} = P_{X_1|U=u} \cdot P_{X_2|U=u} \ldots$. This simple observation greatly simplifies the multi-letter optimization problem: instead of searching for general joint distributions, now we have the further constraint of conditional independence. This directly gives rise to the proof of the optimality of i.i.d. distributions and hence single letter solutions for the $2$ user case, which is the first definitive answer on the general broadcast channels.

The more interesting case is when there are more than 2 receivers. In such cases, i.i.d. distributions simply do not have enough degrees of freedom to be optimal in the tradeoff of more than $2$ linear systems. Instead, one has to design multi-letter product distributions to achieve the optimal. %This is another significant result, since to our knowledge, this is the only case where prove that single letter solutions are not optimal, but still can construct the optimal solution. 
The following example, constructed with the geometric method, illustrate the key ideas. 

\begin{figure}
\centering 
\subfigure[]{
\includegraphics{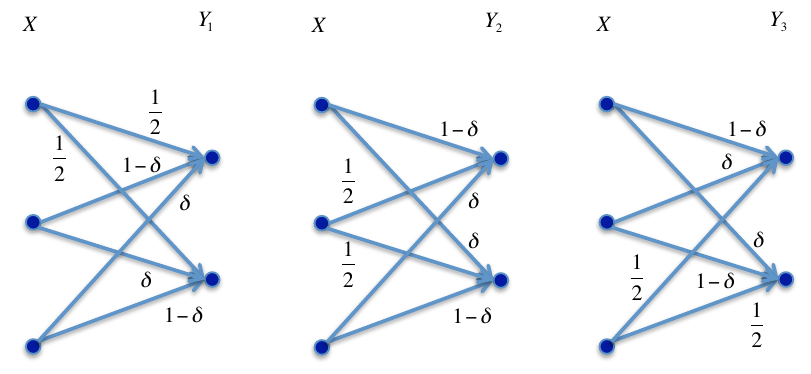} 
\label{fig:Windmill_Channel}
}
\subfigure[]{
\includegraphics{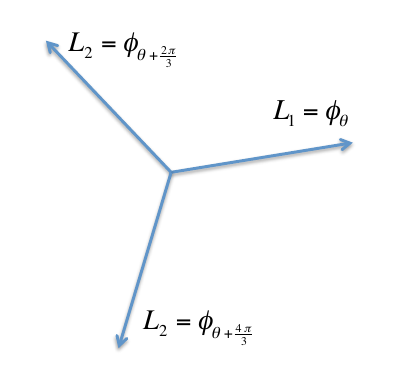}
\label{fig:Windmill}
}
\caption{(a) A ternary input broadcast channel (b) The optimal perturbations over $3$ time slots.}
\end{figure}

\noindent{\bf{Example:}}\\ 
We consider a $3$-user broadcast channel. Let the input alphabet $\cX$ be ternary, so that the perturbation vectors have $2$ dimensions and can be easily visualized. Suppose that the three DTMÕs are rotations of $0$, $2 \pi / 3$, $4 \pi / 3$, respectively, followed (left multiplied) by the projection to the horizontal axis. This corresponds to the ternary input channels as shown in Figure \ref{fig:Windmill_Channel}. Now if we use single-letter inputs, it can be seen that for any $L_{u}$ with $\| L_{u} \|^2 = 1$, $\min \left\{ \| B_1 L_{u} \|^2, \| B_2 L_{u} \|^2, \| B_3 L_{u} \|^2 \right\} \leq 1/4$. The problem here is that no matter what direction $L_{u_0}$ takes, the three output norms are unequal, and the minimum one always limits the performance. Now, if we use $3$-letter input, and denote $\phi_{\theta} = [ \cos \theta , \sin \theta ]^T$, then we can take 
$$P_{X^3| U = u} = (P_X + \epsilon \phi_{\theta}) \otimes (P_X + \epsilon \phi_{\theta + \frac{2 \pi}{3}}) \otimes (P_X + \epsilon \phi_{\theta + \frac{4 \pi}{3}})$$
 for any value of $\theta$, as shown in Figure \ref{fig:Windmill}. Intuitively, this input equalizes the three channels, and gives for all $i = 1, 2, 3$, $\| B^{(3)}_i L^{(3)}_{u} \|^2 = 1/2$, which doubles the information coupling rate. Translating this solution to the coding language, it means that we take turns to feed the common information to each individual user. Note that the solution is not a standard time-sharing input, and hence the performance is strictly out of the convex hull of i.i.d. solutions. One can interpret this input as a repetition of the common message over three time-slots, where the information is modulated along equally rotated vectors. For this reason, we call this example the ``windmill" channel. %While the solution to this toy example is quite intuitive, it is a counter example for the optimality of the MartonÕs region. The most surprising part is that this can be derived simply as the optimal tradeoff between the SVD structures of the DTMÕs.
Additionally, it is easy to see that the construction of the windmill channel can be generalized to the cases of $k > 3$ receivers, where $k$-letter solutions is necessary. 

Note that in this example, we let $U$ be a binary random variable, and in this case, while there are optimal 3-letter solutions, the optimal single-letters do not exist. However, one can in fact take $U$ to be non-binary. For example, let $\cU = \{0,1,2,3,4,5\}$ with $P_U(u) = 1/6$ for all $u$, and let $L_{U=0} = - L_{U=1} = \phi_{\theta}$, $L_{U=2} = - L_{U=3} = \phi_{\theta + \frac{2\pi}{3}}$, and $L_{U=4} = - L_{U=5} = \phi_{\theta + \frac{4\pi}{3}}$, then we can still achieve the information coupling rate 1/2. Thus, there actually exits an optimal single-letter solution with cardinality $|\cU| = 6$. However, when there are $k$ receivers, it requires cardinality $|\cU| = 2k$ for obtaining optimal single-letter solutions. Essentially, this example shows that finding a single perturbation vector with a large image at the outputs of all $3$ channels is difficult. The tension between these $3$ linear systems requires more degrees of freedom in choosing the perturbations, or in other words, the way that common information is modulated. Such more degrees of freedom can be provided either by using multi-letter solutions or have larger cardinality bounds. This effect is not captured by the conventional single-letterization approach.

%Therefore, in general, we can not have optimal single-letter solutions when the cardinality of $\cU$ is bounded by some function of $|\cX|$, but finite-letter optimal solutions still exist, which is essentially what Theorem \ref{thm:BC} says.

Theorem \ref{thm:BC} reduces most of the difficulty of solving the multi-letter optimization problem (\ref{eq:BC_single_public}). The remaining is to find the optimal scaled perturbation $L_{u}$ for the single-letter version of (\ref{eq:BC_single_public}), if the number of receivers $k = 2$, or the $k$-letter version, if $k>2$. These are finite dimensional convex optimization problems \cite{SL04}, which can be readily solved.

We can see that all these information theory problems are solved with essentially the same procedure, and all we need in solving these problems is simple linear algebra. This again, demonstrates the simplicity and uniformity of our approach in dealing with information theory problems.

\section{Conclusion} \label{sec:con}

In this paper, we present the local approximation approach, and show that with this approach, we can handle the issue of single-letterization in information theory by just solving simple linear algebra problems. Moreover, we demonstrate that our approach can be applied to different communication problems with the same procedure, which is a very attractive property. Finally, we provide the geometric insight of the optimal finite-letter solutions in sending the common message to $k>2$ receivers, which also explains why optimal single-letter solutions fail to be existed in these cases.

%\section*{Acknowledgment}

%The authors would like to thank...

\end{document}